# An integration of fast alignment and maximum-likelihood methods for electron subtomogram averaging and classification

Yixiu Zhao[1+], Xiangrui Zeng[1+], Qiang Guo[2], and Min Xu[1*]

[1]School of Computer Science, Carnegie Mellon University, Pittsburgh, USA.
[2]Max Planck Institute for Biochemistry, Martinsried, Germany.
[+]Contributed equally

**Abstract**

**Motivation:** Cellular Electron CryoTomography (CECT) is an emerging 3D imaging technique that visualizes sub-cellular organization of single cells at sub-molecular resolution and in near-native state. CECT captures large numbers of macromolecular complexes of highly diverse structures and abundances. However, the structural complexity and imaging limits complicate the systematic *de novo* structural recovery and recognition of these macromolecular complexes. Efficient and accurate reference-free subtomogram averaging and classification represent the most critical tasks for such analysis. Existing subtomogram alignment based methods are prone to the missing wedge effects and low signal-to-noise ratio (SNR). Moreover, existing maximum-likelihood based methods rely on integration operations, which are in principle computationally infeasible for accurate calculation.

**Results:** Built on existing works, we propose an integrated method, Fast Alignment Maximum Likelihood method (FAML), which uses fast subtomogram alignment to sample sub-optimal rigid transformations. The transformations are then used to approximate integrals for maximum-likelihood update of subtomogram averages through expectation-maximization algorithm. Our tests on simulated and experimental subtomograms showed that, compared to our previously developed fast alignment method (FA), FAML is significantly more robust to noise and missing wedge effects with moderate increases of computation cost. Besides, FAML performs well with significantly fewer input subtomograms when the FA method fails. Therefore, FAML can serve as a key component for improved construction of initial structural models from macromolecules captured by CECT.

# 1 Introduction

Biological pathways rely on the functioning of macromolecular complexes, whose structures and spatial organizations are critical for the function and dysfunction of the pathways. The native structure information of macromolecular complexes has been extremely difficult to obtain due to the limitations of data acquisition techniques. Recent advances in Cellular Electron CryoTomography (CECT) imaging technique enables 3D visualization of subcellular structures at sub-molecular resolution and at near-native state, which makes the extraction of such information possible [22]. However, the imaging limits and high degree of structural complexity make the systematic analysis of a CECT 3D image (i.e. a *tomogram*) highly challenging. The cellular tomograms are normally of very low signal-to-noise ratio (SNR) that few macromolecular complexes can be identified by simple visual inspection. In addition, a tomogram has missing values (i.e. *missing wedge effect*) due to the limited imaging tilt angle range during the data acquisition process, which induces anisotropic resolution. Moreover, the relative size of a macromolecular complex is typically small compared to the image resolution. The abundances of macromolecular structures also vary widely [5]. The structural identification and recovery of macromolecular complexes of low abundance are significantly more difficult than those of high abundance.

Due to the above challenges, the structural recovery of an individual macromolecular complex captured by CECT often requires the inference of its structure (represented by image signals) from large numbers (thousands) of observed subtomograms of identical structures. Such inference is called *subtomogram averaging*. A *subtomogram* is a sub-volume of a tomogram that is likely to contain only one macromolecule. There are two main types of subtomogram averaging methods. The first is through calculating the average of image intensity of multiple aligned subtomograms containing the same structure with the same orientation and displacement (e.g. [10, 3, 32, 1, 8]). Given that the input subtomograms normally contain structures of different orientations and displacements, they need to be aligned to reduce the resolution decrease resulted from orientation and translation difference. Alignment of 3D subtomograms (typical size $\geq 64^3$ voxels) is by nature computationally intensive. Therefore in practice, the alignment based averaging of thousands of subtomograms relies on fast alignment techniques (e.g. [3, 32, 8]) which use approximations to achieve sub-optimal alignment solutions (Section 2.1). Such subtomogram alignment methods (e.g. [32]) were able to achieve three magnitudes of speed increase compared with orientation scanning based exhaustive methods (e.g. [10]). Nevertheless, subtomogram alignment methods are parsimony: they only output a single optimal rigid transformation between a subtomogram and subtomogram average, which is likely to be biased by noise and missing wedge effects. As a result, compared to the maximum-likelihood methods (see the following paragraph), the alignment based subtomogram averaging methods are more prone to noise and missing wedge (Section 3.2).

*Corresponding author email: mxu1@cs.cmu.edu

The second type of averaging methods are maximum-likelihood based (e.g. [27, 7]) (Section 2.3). Compared with alignment based methods, maximum-likelihood methods are, in principle, more robust to noise and to missing wedge effects because the signal at each location is inferred not only from across multiple subtomograms (as in alignment based methods), but also from multiple rigid transforms of each subtomogram through "data augmentation" (Section 2.3). Maximum-likelihood based methods are based on integrating over all rigid transformations. An accurate calculation of such integral in principle requires the exhaustive scanning over a 6D space that parameterizes 3D rigid transformations, which is computationally infeasible.

The macromolecular complexes extracted from cellular tomograms are normally highly heterogeneous. First, crowded cellular environment [6, 12] has macromolecules that adopt different conformations to serve their particular function. They can also dynamically interact with other macromolecules to form different complexes across time. The structural recovery of heterogeneous macromolecules requires separation of the macromolecules into structurally homogeneous sets so that the averaging of each set can more accurately represent the true underlying structures of the set. Such process is called (unsupervised) *subtomogram classification*. The above alignment and maximum-likelihood based subtomogram averaging methods have been extended for simultaneously averaging and classifying the structurally heterogeneous subtomograms (e.g. [3, 32, 9, 27, 7]) by integrating with clustering. The limitations of the averaging methods are also carried to the extended classification tasks. To reduce the heterogeneity of millions of structurally highly diverse macromolecules, we have developed deep learning based unsupervised classification method [33] that can coarsely group subtomograms into more homogeneous clusters without accurate alignment. Clusters of interest can be selected for further analysis.

To complement the above methods, based on existing work, here we propose a new method that integrates the above alignment and maximum-likelihood methods. The new method is termed as integrated Fast Alignment Maximum Likelihood method (FAML). Similar to other subtomogram averaging methods (e.g. [32, 27]), our new method is an expectation-maximization process that iteratively updates the subtomogram averages. However, the updates involve both fast alignment [32] and maximum-likelihood estimation [27]. Specifically, our integrated method consists of three main steps (Algorithm 1): (1) We first calculate a set of the rigid transformations that achieve suboptimal alignments between given subtomograms and subtomogram averages through adapting our previously developed fast alignment method [32] (Section 2.1); (2) We then use these suboptimal rigid transformations to approximate integrals over the entire 6D parametric space of possible 3D rigid transformations (Section 2.2); (3) The approximate integrals are used to update the subtomogram averages through expectation-maximization algorithm similar to [27] (Section 2.3).

Our experiments on simulated and experimental subtomograms show that, compared to our previously developed fast alignment based method (FA) [32, 15], FAML is significantly more robust to noise and missing wedge bias with only a moderate increase in computational costs. FAML also performs well with a low number of input subtomograms when FA fails.

## 2 Methods

An overview of the FAML method is given in Algorithm 1.

**Algorithm 1** Integrated Fast Alignment Maximum Likelihood

1: **procedure** FAML($X = \{X_i, i = 1, \ldots, N\}$)
2:     Initialize model parameters $\Theta = (A, \alpha, \sigma, \xi)$ from the distribution of data $X$ (Section 2.4).
3:     $iter \leftarrow 0$
4:     **for** $iter \leq maxIters$ **do**
5:         $\tilde{\oplus} = \{\oplus_{ik} \leftarrow \text{fastAlign}(X_i, A_k), \forall i = 1, \ldots, N; k = 1, \ldots, K\}$ ▷ Compute suboptimal rigid transformations (Section 2.1)
6:         $\Xi = \{\tilde{\upsilon}(\phi, \oplus_{ik}) \leftarrow \text{voronoiWeights}(\phi, \oplus_{ik}), \forall i = i, \ldots, N; k = 1, \ldots, K; \phi \in \oplus_{ik}\}$ ▷ Compute Voronoi weights (Section 2.2)
7:         $\alpha^{\text{new}} \leftarrow \text{updateAlpha}(X, \Theta, \tilde{\oplus}, \Xi)$ ▷ Equation 11
8:         $\sigma^{\text{new}} \leftarrow \text{updateSigma}(X, \Theta, \tilde{\oplus}, \Xi)$ ▷ Equation 12
9:         $\zeta^{\text{new}} \leftarrow \text{updateXi}(X, \Theta, \tilde{\oplus}, \Xi)$ ▷ Equation 13
10:        $A^{\text{new}} \leftarrow \text{updateA}(X, \Theta, \tilde{\oplus}, \Xi)$ ▷ Equation 10
11:        $\Theta \leftarrow (A^{\text{new}}, \alpha^{\text{new}}, \sigma^{\text{new}}, \zeta^{\text{new}})$ ▷ Update model parameters

## 2.1 Step 1: Calculate suboptimal rigid transformations through fast alignment

Adapted from our previous work [32], we define an alignment score between a subtomogram $X$ and a subtomogram average $A$ as

$$c(\phi^{\text{ro}}, \phi^{\text{tr}}) = \frac{\sum_j w_j^2 X_j \exp(2\pi i \xi_j^\top \phi^{\text{tr}}) \overline{(R_{\phi^{\text{ro}}} A)_j}}{\sqrt{\sum_j w_j^2 \left[ R_{\phi^{\text{ro}}} (A \circ \overline{A}) \right]_j}} \quad (1)$$

The score is essentially a Pearson correlation (up to a constant) restricted only to observed regions of $X$. It is a Fourier space equivalent of a popular alignment score [10]. In Equation 1, $X \in \mathbb{C}^J$ is a $J$ dimensional vector of complex values corresponding to the Fourier representation of a subtomogram. Each element $X_j$ corresponds to the $j$th Fourier component at location $\xi_j \in \mathbb{R}^3$ in Fourier space. Due to the limited tilt angle range in the CECT imaging process (i.e. missing wedge effect), not all components of $X$ can be observed. $X$ is divided into observed and missing components, indicated by a $J$-dimensional indicator vector $\mathbf{w}$ (i.e. missing wedge mask), such that $w_j = 1$ if $X_j$ is observed, and $w_j = 0$ if $X_j$ is unobserved. Similarly, the subtomogram average $A \in \mathbb{C}^J$ is also a $J$ dimensional vector of complex values. $R_{\phi^{\text{ro}}}$ is the 3D rotation operator parameterized by three Euler angles $\phi^{\text{rot}} = (q_\alpha, q_\beta, q_\gamma)$ in ZYZ convention. $\phi^{\text{tr}} = (q_x, q_y, q_z)^\top \in \mathbb{R}^3$ is a vector that corresponds to 3D real-space translation of $A$. $\circ$ denotes entry-wise product.

The alignment of $X$ and $A$ is a process of finding the optimal rotation $\phi^{\text{ro}}$ and translation $\phi^{\text{tr}}$ that maximizes $\mathscr{R}(c)$, which is the real part of $c$. The direct optimization of $\mathscr{R}(c)$ requires scanning through all sampled rotation and use Fast Fourier Transform (FFT) to exhaustively search through all sampled translation for each rotation. Such exhaustive scanning based optimization is highly computationally intensive [32] and has very limited scalability. Therefore, we compute a set of suboptimal rigid transforms using a translation-invariant upper-bound $\tilde{c}(\phi^{\text{ro}}) = |c(\phi^{\text{ro}}, \phi^{\text{tr}})|$ of $\mathscr{R}(c)$. Specifically, we express $\tilde{c}$ as a fraction of two rotational correlation functions $\tilde{c}_0$ and $\tilde{c}_1$:

$$\tilde{c}(\phi^{\text{ro}}) = \frac{\tilde{c}_0(\phi^{\text{ro}})}{\tilde{c}_1(\phi^{\text{ro}})} = \frac{\sum_j (p_0)_j \overline{(R_{\phi^{\text{ro}}} q_0)_j}}{\sqrt{\sum_j (p_1)_j \overline{(R_{\phi^{\text{ro}}} q_1)_j}}} \quad (2)$$

, where $p_0 = \mathbf{w} \circ \mathbf{w} \circ X$, $q_0 = A$, $p_1 = \mathbf{w} \circ \mathbf{w}$, $q_1 = A \circ \overline{A}$. After representing $p$ and $q$ using spherical harmonics expansion, these rotational correlation functions are efficiently and simultaneously calculated over all rotations [20] using the Fast Fourier Transform (FFT) through re-parameterization. The set $\oplus^{\text{ro}}$ of suboptimal rotations is then obtained according to the local maxima of $\tilde{c}$ in the 3D parameter space of $\phi^{\text{ro}}$. The corresponding suboptimal translation $\phi^{\text{tr}}(\phi^{\text{ro}})$ for each $\phi^{\text{ro}} \in \oplus^{\text{ro}}$ is then calculated using FFT in a similar way as in [32, 10]. We denote $\oplus := \{(\phi^{\text{ro}}, \phi^{\text{tr}}(\phi^{\text{ro}})), \forall \phi^{\text{ro}} \in \oplus^{\text{ro}}\}$ as the final set of suboptimal rigid transformations. In practice, the size of $\oplus$ is normally smaller than 50.

## 2.2 Step 2: Approximate integration by summation over suboptimal rigid transformations

In the maximum-likelihood based subtomogram averaging methods (e.g. [27]) (Section 2.3), the updating of subtomogram averages are based on the calculation of the integrals of the following form

$$\int_\phi f(\phi, X, A) \, d\phi \quad (3)$$

for a function $f$ of rigid transformation $\phi$, subtomogram $X$, and subtomogram average $A$. However, the accurate calculation of Equation 3 in principle requires exhaustively scanning through all rigid transformations in a 6D parameter space that consists both rotational $\phi^{\text{ro}}$ and translational $\phi^{\text{tr}}$ parameters. Such exhaustive scanning is computationally infeasible (Section 3.4). In this paper, we approximate the integral in Equation 3 over all rigid transformations with a small set $\oplus$ of sub-optimal transformations obtained from Section 2.1:

$$\int_{\phi} f(\phi, X, A)\, d\phi \approx \sum_{\phi \in \oplus} f(\phi, X, A)\, \tilde{v}(\phi, \oplus), \tag{4}$$

where $\tilde{v}(\phi, \oplus) := \frac{|v(\phi, \oplus)|}{\sum_{\phi' \in \oplus} |v(\phi', \oplus)|}$ is the normalized hypervolume of $\phi$, and $|v(\phi, \oplus)|$ is the hypervolume of the Voronoi region $v(\phi, \oplus)$ of $\phi \in \oplus$ on the manifold $\Upsilon \subset \mathbb{R}^6$ that parameterize all rigid transformations. For those $\phi \in \oplus$, the $\sum_{j=1}^{J_i^o} ||(R_\phi A_k)_j - X_{ij}^o||^2$ term in Equation 15 tend to be small. Therefore in Equation 14 the corresponding $e(\phi, k)$ tend to have large contribution to the calculation of the probability.

Each rigid transformation corresponds to a point in $\Upsilon$. To calculate the hypervolume $|v(\phi, \oplus)|$, we use a Monte-Carlo sampling method (Figure 1) that is similar to [2]. For the three position coordinates $(q_x, q_y, q_z)$ of $\phi$, we set the sampling boundaries to the minimum and maximum values among the set of points along each axis. For the three rotational coordinates $(q_\alpha, q_\beta, q_\gamma)$ of $\phi$, we treat them as independent coordinates that wrap around after $2\pi$. We randomly sample a point $\phi' \in \Upsilon$, and compute its nearest neighbor

$$\phi^*(\phi') = \underset{\phi \in \oplus}{\arg\min}\ \tau(\phi', \phi). \tag{5}$$

In such case, $\phi'$ will belong to $v(\phi^*(\phi'), \oplus)$. After sampling a set $\oplus'$ of a large number (e.g. 10000) of points in $\Upsilon$ uniformly, we approximate

$$|v(\phi, \oplus)| \approx \frac{|\{\phi' \in \oplus' \mid \phi^*(\phi') = \phi\}|}{|\oplus'|},\ \forall \phi \in \oplus. \tag{6}$$

For any two points $\phi_1, \phi_2 \in \Upsilon$, we define their distance as $\tau(\phi_1, \phi_2) = \sqrt{||\mathbf{I} - \mathbf{R}_1^\top \mathbf{R}_2||_F^2 + \alpha^{\text{trd}} ||\phi_1^{\text{tr}} - \phi_2^{\text{tr}}||_2^2}$, where $\mathbf{R}$ is the corresponding rotation matrix of $\phi^{\text{rot}}$, $||\cdot||_F$ is the Frobenius norm [19], and $\alpha^{\text{trd}}$ is a coefficient parameter used to balance the value scales between rotation and translation. In our experiments (Section 3), we set $\alpha^{\text{trd}} = 1$ for simplicity.

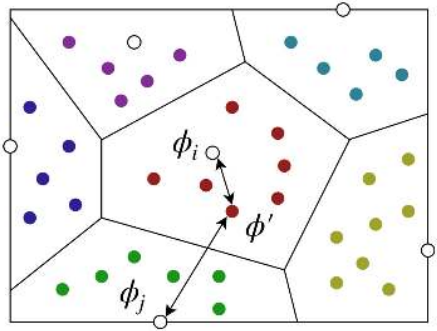

Figure 1: The basic idea of hypervolume calculation. The white dots are the suboptimal rigid transformations $\phi \in \oplus$ obtained by fast alignment (Section 2.1), and the colored dots are the sampled points $\phi' \in \oplus'$. The Voronoi region $v(\phi_i, \oplus)$ is defined as the set of all points $\phi' \in \Upsilon$ such that it is closer to $\phi_i \in \oplus$ than any $\phi_j \in \oplus$ when $i \neq j$ under the distance metric $\tau$. The number of points in a Voronoi region becomes a good estimation of its hypervolume when a large number of points are sampled.

### 2.3 Step 3: Maximum-likelihood based updating of subtomogram averages using expectation-maximization

We follow the data model and notations defined in [27]:

$$X_i = R_{\phi_i} A_{\kappa_i} + G_i\ \forall i = 1, \ldots, N, \tag{7}$$

where

- $N$ is total number of input subtomograms

- $X_i \in \mathbb{C}^J$ is the $i$th subtomogram in the form of a $J$-dimensional vector of complex values $(X_i)_j$ (or $X_{ij}$ in short), which is divided into a vector $X_i^o$ of observed components $X_{ij}^o$ and a vector $X_i^m$ missing components $X_{ij}^m$.

- The observed and missing components of $X_i$ are formalized by defining a $J$-dimensional missing data indicator vector $\mathbf{w}_i \in \{0,1\}^J$, such that $(\mathbf{w}_i)_j = 1$ if $(X_i)_j$ is observed, and $(\mathbf{w}_i)_j = 0$ if $(X_i)_j$ is missing.

- $A_{\kappa_i}$ is one of $K$ unknown 3D structures represented by subtomogram averages in Fourier space. $A_1, \ldots, A_K \in \mathbb{C}^J$. These are the objects to estimate from the data. The data model is used for subtomogram averaging when setting $K = 1$, and subtomogram classification and averaging when setting $K > 1$.

- $\kappa_i \in \{1, 2, \ldots, K\}$ is an unknown, random integer, indicating which of the unknown structures corresponded to $X_i$.

- $R_{\phi_i}$ is a rigid transformation operator which maps the unknown structure $A_{\kappa_i}$ onto $X_i$. The actual transformation $\phi_i$ for particle $X_i$ are unknown. Same as in Section 2.1, this transformation is parameterized by $\phi$, a 6D vector (corresponding to three Euler angles $\phi^{\text{rot}} = (q_\alpha, q_\beta, q_\gamma)$ in ZYZ convention, and three real-space translation coordinates, $\phi^{\text{tr}} = (q_x, q_y, q_z)$); In such case, the rigid transformation operator $R_\phi$ is decomposed into a combination of rotation and translation operators $R_\phi := R_{\phi^{\text{tr}}} R_{\phi^{\text{ro}}}$.

- $G_i \in \mathbb{C}^J$ is a $J$-dimensional vector of unknown, independent Gaussian noise with zero mean and unknown standard deviation $\sigma$.

Given this model, the complete data set corresponds to

$$(X_i^o, X_i^m, \phi_i, \kappa_i) \ \forall \ i = 1, \ldots, N \tag{8}$$

Subtomogram classification and averaging based on the data model in Equation 7 can be treated as an extension of model-based clustering process [11] that aims to find parameters that maximize the approximate log-joint probability of observing the entire set of observed data with the data model defined in Equation 7 [27]:

$$L(\Theta) = \sum_{i=1}^{N} \log \sum_{k=1}^{K} \int_\phi \int_{M_i} P(X_i^o | k, \phi, X_i^m, \Theta) P(k, \phi, X_i^m | \Theta) \, d\phi \, dM_i, \tag{9}$$

where the probabilities are modeled in the same way as in [27]. $\int_{M_i} dM_i$ is a shorthand notation for the integrals for every missing Fourier component in $X_i^m$ [27].

In this paper, we maximize $L(\Theta)$ by Expectation-Maximization through similar derivation as [27], but with approximate integrals according to Section 2.2. The derived equations for updating the averages and other parameters are as follows:

$$A_{kj}^{\text{new}} = \frac{1}{N \alpha_k^{\text{new}}} \sum_{i=1}^{N} \sum_{\phi \in \oplus_{ik}} P(k, \phi | X_i^o, \Theta) \left\{ [R_\phi^{-1} w_i]_j (R_\phi^{-1} X_i^o)_j + [1 - (R_\phi^{-1} w_i)_j] A_{kj} \right\} \tilde{v}(\phi, \oplus_{ik}) \tag{10}$$

$$\alpha_k^{\text{new}} = \frac{1}{N} \sum_{i=1}^{N} \sum_{\phi \in \oplus_{ik}} P(k, \phi | X_i^o, \Theta) \, \tilde{v}(\phi, \oplus_{ik}) \tag{11}$$

$$(\sigma^{\text{new}})^2 = \frac{1}{NJ} \sum_{i=1}^{N} \sum_{j=1}^{J} \sum_{k=1}^{K} \sum_{\phi \in \oplus_{ik}} P(k, \phi | X_i^o, \Theta) \left\{ w_{ij} \|(R_\phi A_k)_j - X_{ij}^o\|^2 + (1 - w_{ij})(\sigma)^2 \right\} \tilde{v}(\phi, \oplus_{ik}) \tag{12}$$

$$(\zeta^{\text{new}})^2 = \frac{1}{3N} \sum_{i=1}^{N} \sum_{k=1}^{K} \sum_{\phi \in \oplus_{ik}} P(k, \phi | X_i^o, \Theta) \left\{ q_x^2 + q_y^2 + q_z^2 \right\} \tilde{v}(\phi, \oplus_{ik}) \tag{13}$$

$$P(k, \phi | X_i^o, \Theta) = \frac{e(\phi, k)}{\sum_{k'} \sum_{\phi' \in \oplus_{ik'}} e(\phi', k') \, \tilde{v}(\phi', \oplus_{ik'})} \tag{14}$$

$$e(\phi,k) := \alpha_k \exp\left\{\frac{\sum_{j=1}^{J_i^o}||(R_\phi A_k)_j - X_{ij}^o||^2}{-2\sigma^2} + \frac{q_x^2 + q_y^2 + q_z^2}{-2\zeta^2}\right\} \quad (15)$$

Optionally, regularization of the similarities between averages can also be applied in a similar way as in [27].

## 2.4 Parameter initialization

The way we initialize the model parameters is as follows. We set the $\alpha_i$ for every class to be equal to $1/K$. We divide all subtomograms evenly into $K$ sets at random and let the average of each set be the class average $A_i$. The initial value of $\sigma^2$ is obtained by picking a random subtomogram and class average and computing the square voxel intensity difference averaged over all observed parts, and $\zeta$ is initially set to be equal to the size of the image.

## 2.5 Implementation details

A modified version of the Tomominer library [15] was used for fast alignment, 3D rigid transformation, and other processing. EMAN2 [16] library was used for constructing simulated subtomograms. The methods were parallelized on multiple CPU cores. The tests were performed on two computers. The first computer has two Intel Xeon E5-2687W CPUs at 3.0GHz frequency and 256GB memory, allowing simultaneous running 48 parallel processes. The second computer has one Intel Core i7-6800K CPU at 3.4GHz frequency and 128GB memory, allowing simultaneous running 12 parallel processes. The isosurfaces and atomic models were plotted using UCSF Chimera [26]. In all tests, both FAML and FA methods were executed for 20 iterations and converged within 20 iterations.

# 3 Results

## 3.1 Generation of realistically simulated subtomograms

In order to assess the performance of the FAML method, we simulated realistic subtomograms by mimicking the tomogram reconstruction process as described previously [10, 4, 24]. Missing wedge, image noise, and electron optical factors, including the Modulation Transfer Function (MTF) and the Contrast Transfer Function (CTF), were properly included. Electron optical density of macromolecular complexes was set to be proportional to the electrostatic potential. Volume electron density maps were generated by the Situs [30] PDB2VOL program, which was used to simulate electron micrograph images through a sequence of tilt-angles. Random noise was added to the images [10] to reach the target SNR levels, which were similar to the SNRs estimated from experimental data (Section 3.2.2). Electron micrograph images were convolved with the MTF and CTF to produce electron optical effects [24, 13]. Data acquisition parameters in the simulation were determined by the experimental data acquisition parameters in Section 3.2.2, with spherical aberration of 2.7 mm, defocus of -6$\mu$ m, and voltage of 300 kV. The MTF is defined as $\text{sinc}(\pi\omega/2)$, where $\omega$ is the fraction of the Nyquist frequency, corresponding to a detector [23]. To reconstruct the simulated subtomogram from the tilt series, a direct Fourier inversion reconstruction algorithm (from the EMAN2 library [16]) was applied.

To determine the SNR of experimental subtomograms, we measured the SNR of the selected 859 ribosome subtomograms from a tomogram of primary rat neuron culture (Section 3.2.2). We selected 1000 random pairs of subtomograms that were already aligned to the corresponding ribosome template (PDB ID: 5T2C). The SNR of each subtomograms pairs was computed according to [14]. The mean SNR is 0.01035. We measured the SNR of TMV subtomograms (Section 3.2.3) in a similar way by aligning them to their FAML average. The mean SNR is 0.002313. The measured SNRs serve as a reference range to determine the SNR of simulated subtomograms. All simulated subtomograms are of size $64^3$ with voxel size 0.6 nm and resolution 0.6 nm.

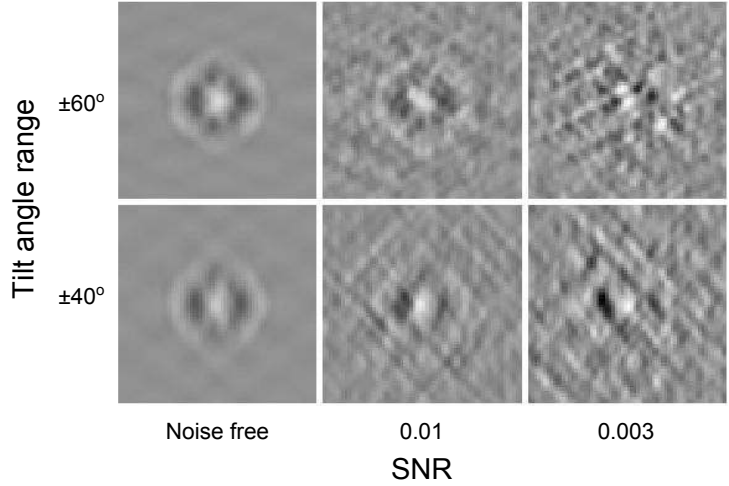

Figure 2: Center slices (x-z plane) of simulated subtomograms of specified level of SNRs and tilt angle ranges.

Figure 2 shows central slices of simulated GroEL (PDB ID: 1KP8) subtomograms (size: $64^3$) of different level of SNRs and tilt angles. Compared with noise-free subtomograms (i.e. templates), subtomograms of lower SNR and smaller tilt angle range show more distortions.

### 3.2 Reference-free averaging tests

When we assume all input subtomograms contain the same structure with random orientations and displacements, we choose $K = 1$. FAML is used for refining a single average, known as reference-free averaging.

#### 3.2.1 Averaging of simulated GroEL subtomograms

Due to the crowded cellular environments and imaging limits, CECT data is usually of low SNR. Low SNR is a major challenge for reference-free subtomogram averaging. To test the performance of the FAML averaging with respect to a high noise level, we chose a low SNR 0.003, and simulated 100 GroEL (PDB ID: 1KP8) at that SNR level with tilt angle range $\pm 60°$. All 100 GroEL structures were randomly rotated and translated before constructing the simulated subtomograms.

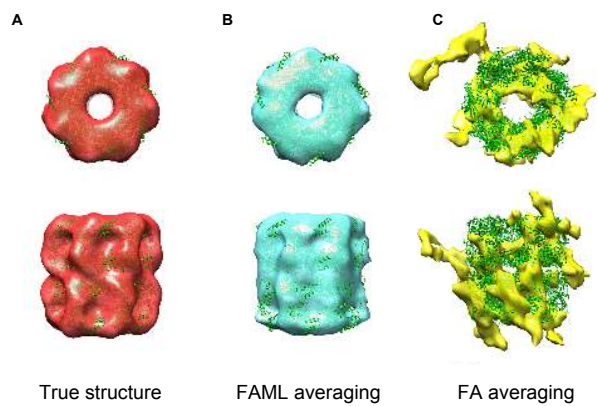

Figure 3: Averaging of low SNR simulation GroEL subtomograms: (A) Isosurface of true GroEL structure (PDB ID: 1KP8, filtered at 0.6 nm resolution) with fitted atomic model. (B) FAML subtomogram average with fitted atomic model ($r = 0.77$). (C) FA subtomogram average with fitted atomic model ($r = 0.19$).

The averaging results were plotted with fitted atomic model alongside a true GroEL structure (Figure 3). The fitted atomic model with FAML GroEL average achieved cross-correlation coefficient of 0.77 whereas the fitted atomic model with FA GroEL average achieved cross-correlation coefficient of 0.19. Figure 3C showed that at such a low SNR level, the FA method failed to recover the GroEL structure, resulting in a subtomogram average of a collection of 'torn pieces'. Figure 3B showed that FAML method successfully recovered the GroEL structure. The top view (Figure 3B top) showed that the seven-fold rotational symmetry of GroEL was recovered. The advantage of FAML over FA on low SNR experimental subtomograms was further demonstrated in Section 3.2.2.

Another CECT imaging distortion is the missing wedge effect. Many tomograms are processed at a small tilt angle range such as $\pm 40°$ or $\pm 50°$ to prevent excessive electron beam damage to the specimen. As the same structure may adopt different orientations inside a subtomogram, the missing wedge bias could be partly compensated and corrected by aligning and averaging multiple identical structures of different orientations during the structural recovery process. However, in some cases, such as small numbers of input subtomograms, small tilt angle ranges (i.e. large missing wedge angles), and having preferred orientations, correcting the missing wedge bias in the averaging process is substantially more challenging. In fact, a recent study shows that having preferred orientation is often a problem in single particle cryo-electron microscopy imaging [17].

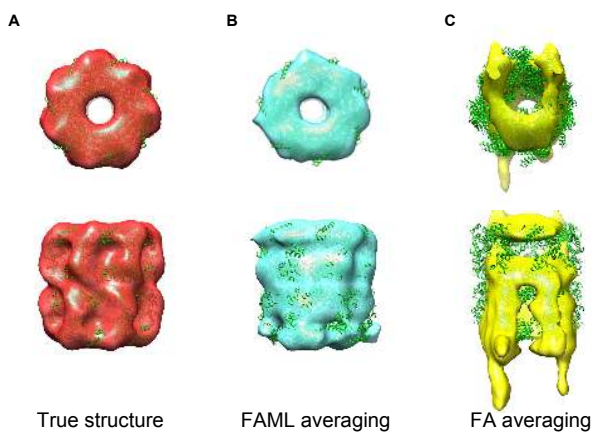

Figure 4: Averaging of simulated GroEL subtomograms with small tilt angle range: (A) Isosurface of true GroEL structure (PDB ID: 1KP8, filtered at 0.6 nm resolution) with fitted atomic model. (B) FAML subtomogram average with fitted atomic model ($r = 0.78$). (C) FA subtomogram average with fitted atomic model ($r = 0.49$).

To test the performance of the FAML method on reference-free averaging of subtomograms with the above limits, we simulated a small number of 50 GroEL (PDB ID: 1KP8) subtomograms, at SNR level 0.01 with tilt angle range $\pm 40°$. Preferred orientation was also simulated by only allowing the structure to rotate about the Y axis. The principal axis of the structure is constrained to be parallel to the Y axis. All the GroEL structures were randomly rotated with constraints and translated before constructing simulated subtomograms.

The averaging results were plotted with fitted atomic model alongside a true GroEL structure (Figure 4). The fitted atomic model with FAML GroEL average achieved cross-correlation coefficient of 0.78 whereas the fitted atomic model with FA GroEL average achieved cross-correlation coefficient of only 0.49. Figure 4C showed that the FA method is heavily biased by the missing wedge effect. The averaged structure is distorted and elongated along the Y direction with a sizable missing region along the Z direction. In comparison, the FAML method fully corrected the missing wedge effect (Figure 4B). No missing regions nor significant distortions are observable from the FAML average. The top view (Figure 4B top) showed that FAML method recovered the GroEL structure with the seven-fold rotational symmetry. The advantage of FAML over FA on reducing missing wedge bias was further demonstrated on the experimental TMV subtomograms in Section 3.2.3.

### 3.2.2 Averaging of experimental ribosome subtomograms extracted from a tomogram of primary rat neuron culture

Reference-free averaging was also tested on a dataset of 859 ribosome subtomograms extracted and purified from a tomogram of primary rat neuron culture [18]. The tomogram was captured with a tilt angle range of $-50°$ to $+70°$.

It was then binned twice to a voxel size of 1.368 nm. 58549 subtomograms of size $40^3$ were extracted from the tomogram using the Difference of Gaussian particle picking method [25]. The extracted subtomograms are highly heterogeneous. Therefore, we used a convolutional autoencoder [33] to perform unsupervised clustering of the extracted subtomograms and selected only the clusters with large globular features because they are more likely to be ribosomes. This filtering process selected about 10 % subtomograms for further analysis. Template search [15] was applied to identify the top 1000 subtomograms with high structural correlation to the ribosome template. We manually inspected the 1000 subtomograms, and filtered out 141 of them which contained obvious non-ribosome structure such as fiducial.

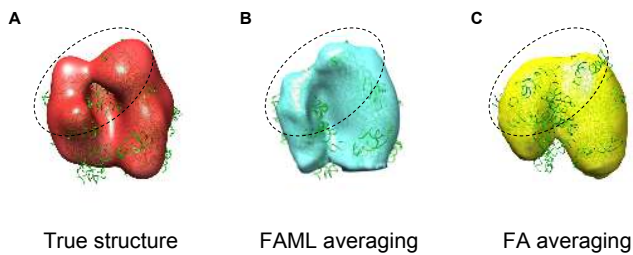

Figure 5: Averaging of experimental ribosome subtomograms (circled regions show that FAML recovers more structural details): (A) Isosurface of true ribosome structure (PDB ID: 5T2C, filtered at 10 nm resolution) with fitted atomic model. (B) FAML subtomogram average with fitted atomic model ($r$ = 0.61). (C) FA subtomogram average with fitted atomic model ($r$ = 0.66).

Both FAML and FA method were tested on this ribosome subtomogram dataset. The averaging results were plotted with fitted atomic models alongside a true ribosome structure filtered at low resolution (10 nm). The fitted atomic model with FAML ribosome average achieved cross-correlation coefficient of 0.61 whereas the fitted atomic model with FA ribosome average achieved cross-correlation coefficient of 0.66. Figure 5C showed that FA subtomogram average converged to a general shape resembling a ribosome structure consisting of the 40S and 60S subunits with a major glove feature in between. Although the cross-correlation coefficient for FA average is slightly higher, the finer structural details (circled region), such as those connecting the two subunits, were lost as compared to the true structure. FAML method, on the other hand, recovered not only the general shape of a ribosome with two subunits but also with significantly more structural details of both subunits and their connection.

### 3.2.3 Averaging of experimental tobacco mosaic virus subtomograms

The performance of FAML in correcting missing wedge effects was further tested using a subtomogram dataset of tobacco mosaic virus (TMV), a type of helical virus [21]. The dataset consists of 2742 TMV subtomograms of size $128^3$. They were two times binned to size $64^3$. The tilt angle range is $\pm 60°$ and the voxel size is 0.54 nm after binning.

Without taking advantage of rotational symmetry information, the FA subtomogram averaging resulted in large missing regions (Figure 6B bottom part) and appears to be a stack of ring structures rather than a single helical structure (Figure 6D). The top and bottom view regions of the FA averages are significantly distorted (Figure 6B). Compared to the FA average, the FAML average is significantly more similar to the known helical structure of TMV, although the parts that are located at top and bottom views are not perfectly smooth (Figure 6A). No significant missing regions are observable and a symmetric helical structure was roughly recovered (Figure 6C).

It is known the TMV has seventeen-fold symmetry. We measured the symmetry of FA and FAML averages. The symmetry was measured by the pair-wise correlation between the structure and its rotation along the principal axis with an angle corresponding to the seventeen-fold symmetry. For each average, seventeen rotated structures were generated and the average pairwise correlation was computed.

FAML achieved an average pairwise correlation of 0.47. FA achieved an average pairwise correlation of 0.25. Therefore, the FAML average recovered better TMV symmetric features.

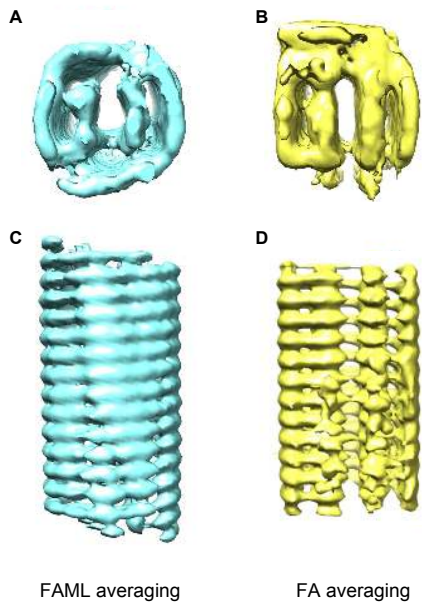

FAML averaging      FA averaging

Figure 6: Averaging of experimental TMV subtomograms: (A) Isosurface of FAML subtomogram average (symmetry: 0.47). (B) FA subtomogram average (symmetry: 0.25).

### 3.3 Reference-free classification and averaging tests

Frequently, subtomogram datasets contain heterogeneous structures. Simple averaging will result in recovering a mixed structure. In such case, unsupervised classification should be performed simultaneously with the subtomogram averaging process to recover multiple averages of different structures. Structural recovery accuracy, as well as classification accuracy, are both important for these tasks.

#### 3.3.1 Classification and averaging of simulated GroEL and ribosome subtomograms

To test the performance of FAML on reference-free classification and averaging tasks, we simulated 100 GroEL (PDB ID: 1KP8) and 100 ribosome (PDB ID: 4V4A) subtomograms at SNR level 0.01 with tilt angle range $\pm 60°$. All 200 structures were randomly rotated and translated inside the subtomogram.

This dataset of 200 subtomograms was classified and averaged by both FAML and FA methods with $K = 2$. FAML method successfully classified the 200 subtomogram into 2 classes, of which class 1 contains 100 GroEL subtomograms and class 2 contains 100 ribosome subtomograms. No subtomogram was misclassified. The averaging results of the two classes were plotted alongside a true GroEL structure and a true ribosome structure (Figure 7). The characteristic seven-fold symmetry feature of the GroEL structure was successfully recovered (Figure 7B top). The ribosome average resembles the true structure in terms of structural details to a good extent (Figure 7E).

By contrast, FA method classified the 200 subtomogram into 2 classes, of which class 1 contains 98 GroEL subtomograms and class 2 contains 100 ribosome subtomograms and 2 GroEL subtomograms. 2 subtomograms were misclassified. The averaging results of the two classes were plotted alongside a true GroEL structure and a true ribosome structure (Figure 7). The seven-fold symmetry of the GroEL structure was successfully recovered because the classified GroEL class contains 98 homogeneous GroEL structures (Figure 7C top). However, additional structures were falsely generated in the top region of ribosome average (Figure 7F). This is mainly due to the two GroEL subtomograms mixed into the ribosome subtomogram class.

The fitted atomic model with FAML GroEL and ribosome averages achieved cross-correlation coefficients of 0.88 and 0.85, respectively whereas the fitted atomic model with FA GroEL and ribosome averages achieved cross-correlation coefficients of 0.84 and 0.75, respectively.

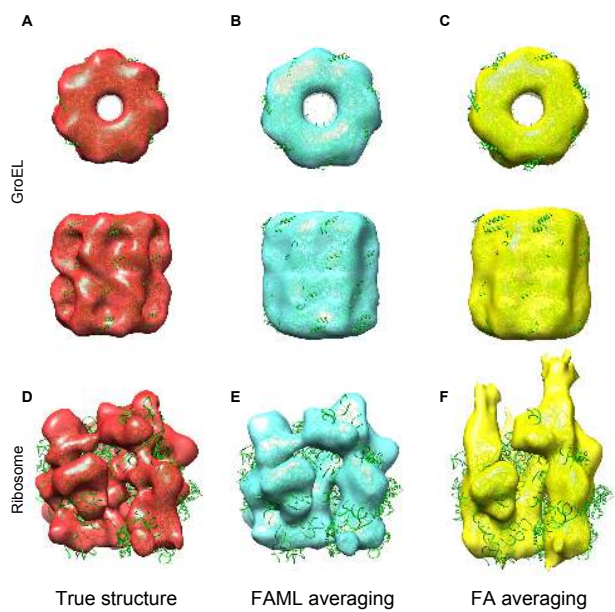

Figure 7: Classification and averaging of simulated GroEL and ribosome subtomograms: (A) Isosurface of true GroEL structure (PDB ID: 1KP8, filtered at 0.6 nm). (B) FAML average of GroEL subtomogram ($r = 0.88$). (C) FA average of GroEL subtomogram ($r = 0.84$). (D) True ribosome structure (PDB ID: 4V4A, filtered at 0.6 nm). (E) FAML average of ribosome subtomogram ($r = 0.85$). (F) FA average of ribosome subtomogram ($r = 0.75$).

We measured the classification accuracy in terms of F1 score, which is the harmonic mean of precision and recall. Overall, the FAML method achieved an average F1 score of 1 and the FA method achieved an average F1 score of 0.99. The FAML method outperforms the FA method regarding both classification accuracy and structural recovery accuracy.

### 3.3.2 Averaging and classification of experimental capped proteasome and TRiC subtomograms extracted from a tomogram of rat neuron with expression of poly-GA aggregate

Furthermore, reference-free classification and averaging were tested on a dataset consisting of 125 TCP-1 ring complex (TRiC) subtomograms and 200 capped proteasome subtomograms extracted from a tomogram of rat neuron with expression of poly-GA aggregate [18]. All subtomorgams were two times binned to size $40^3$ (voxel size: 1.368 nm). The tilt angle range was $-50°$ to $+70°$.

The reference-free classification and averaging tasks were substantially more challenging due to the small number of input subtomograms. The TRiC&proteasome dataset was classified and averaged by both FAML and FA methods with $K = 2$.

The averaging results of the two classes were plotted alongside a true TRiC structure and a true capped proteasome structure (Figure 8). The fitted atomic model with FAML TRiC and capped proteasome averages achieved cross-correlation coefficients of 0.41 and 0.45, respectively whereas the fitted atomic model with FA TRiC and capped proteasome averages achieved cross-correlation coefficients of 0.08 and 0.16, respectively.

FAML method classified the 325 subtomogram into 2 classes, of which class 1 contains 91 TRiC subtomograms and 6 capped proteasome subtomograms, and class 2 contains 194 capped proteasome subtomograms and 34 TRiC subtomograms. 40 subtomograms were misclassified. FAML recovered the spherical shape of TRiC to a similar size (Figure 8B). The capped proteasome average resembles the true structure in terms of its cylindrical shape and the cap on the top (Figure 8E).

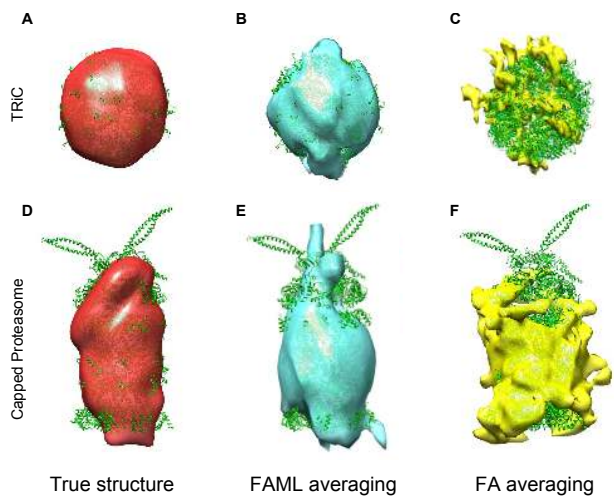

Figure 8: Averaging and classification of experimental capped proteasome and TRiC subtomograms: (A) True TRiC structure (PDB ID: 4V94, filtered at 6 nm). (B) FAML subtomogram average ($r = 0.41$). (C) FA subtomogram average ($r = 0.08$). (D) True capped proteasome structure (PDB ID: 5MPA, filtered at 6 nm). (E) FAML subtomogram average ($r = 0.45$). (F) FA subtomogram average ($r = 0.16$).

On the other hand, the FA method classified the 325 subtomogram into 2 classes, of which class 1 contains 10 TRiC subtomograms, and class 2 contains 200 capped subtomograms and 115 TRiC subtomograms. 115 subtomograms were misclassified. The averaging results of the two classes were plotted alongside a true TRiC structure and a true capped proteasome structure (Figure 8). The spherical shape of the TRiC structure was not recovered due to the low number of TRiC subtomograms classified in class 1 . The capped proteasome structure was also not correctly recovered mainly due to the high number of TRiC subtomograms misclassified to class 2 ( Figure 8F).

Overall, the FAML method achieved an average F1 score of 0.863 and the FA method achieved an average F1 score of 0.462. The FAML method significantly outperformed the FA method in terms of accuracy.

### 3.3.3 Averaging and classification of experimental GroEL and GroEL-GroES subtomograms

We tested the performance of FAML method on classifying and averaging subtomograms with high structural similarity. Reference-free averaging and classification were tested using a dataset of experimental GroEL and GroEL-GroES subtomograms captured in [10]. The dataset consists of 780 subtomograms belonging to two class: GroEL and GroEL-GroES. To show that the FAML method can achieve successful averaging and classification with a small number of input subtomograms, we substantially decreased the size of the GroEL/GroEL-GroES dataset by randomly selecting 400 subtomograms. All the 400 subtomograms are of size $32^3$ with voxel size 1.2 nm and tilt angle range $\pm 65°$.

Both FAML and FA methods were applied to the selected subtomograms. The averaging results of the two classes were plotted alongside a true GroEL structure and a true GroEL-GroES structure. (Figure 9). Though the FA method was tested previously on the original dataset of 780 subtomograms and successfully recovered the GroEL and GroEL-GroES structure [15], when decreasing the input subtomogram number to 400, the FA method could not fully recover either the GroEL structure (Figure 9C) or the GroEL-GroES structure (Figure 9F). Both structures are heavily distorted compared to the true structures.

By contrast, the FAML method recovered both GroEL (Figure 9B) and GroEL-GroES (Figure 9E) structures as compared to the true structures regarding their size and symmetric shape. The averaged GroEL-GroES structure can be distinguished from the averaged GroEL structure by its characteristic enlarged chamber at its top (Figure 9E bottom).

The fitted atomic model with FAML GroEL and GroEL-GroES averages achieved cross-correlation coefficient of 0.87 and 0.78, respectively whereas the fitted atomic model with FA GroEL and GroEL-GroES averages achieved cross-correlation coefficients of 0.40 and 0.24, respectively. The previously reported performance for method [27] on the whole 780 subtomograms is 0.88 and 0.81 for GroEL and GroEL-GroES averages, which is only slightly higher than ours obtained from a significantly smaller number of only 400 subtomograms.

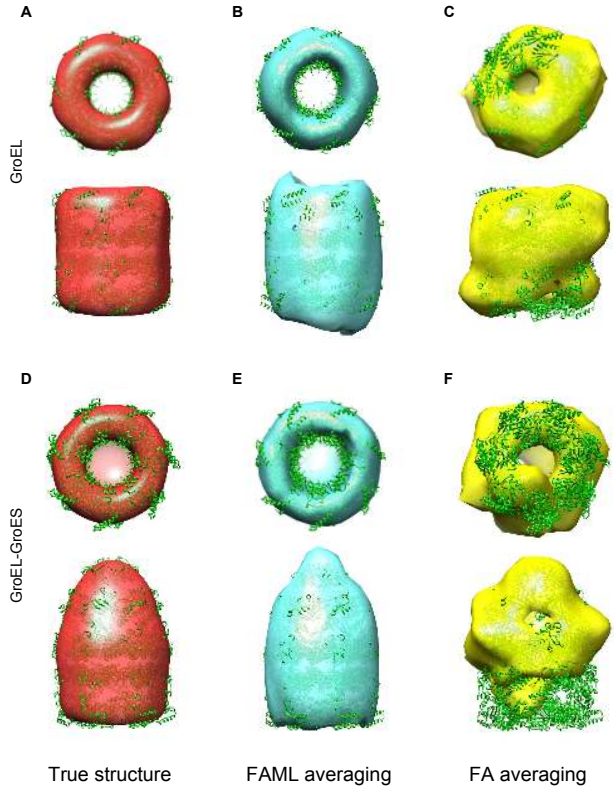

Figure 9: Averaging and classification of experimental GroEL and GroEL-GroES subtomograms: (A) Isosurface of true GroEL structure (PDB ID: 1KP8, filtered at 6 nm) . (B) FAML subtomogram average ($r = 0.87$). (C) FA subtomogram average. ($r = 0.40$) (D) True GroEL-GroES structure (PDB ID: 2C7C, filtered at 6nm). (E) FAML subtomogram average ($r = 0.78$). (F) FA subtomogram average ($r = 0.24$).

Therefore, the FAML method significantly outperformed the FA method in classification and averaging of similar structure with a substantially smaller number of input subtomograms. The FAML method achieved comparable averaging results to method [27] on substantially fewer input subtomograms.

## 3.4 Computation time analysis

Similar to FA, the computational cost for all the components of FAML scales linearly with the number of voxels in the input subtomograms. The cost for the maximum-likelihood update functions is linear to the size of sampled rigid transformations $\oplus$ taken to approximate the integrals. To give an estimation of how our method compares with those with exhaustive integration, we performed time profiling during execution of FAML. As we can see in the graph (Figure 10), the steps for maximum-likelihood update in FAML cost 66% of the time. Using a uniform grid sampling to obtain $\oplus$ instead of using fast alignment and aim for no more than a 10x increase of computational cost, we can at most afford to increase the size of $\oplus$ by a factor of 20. Given that our method normally uses at most 50 samples for each integration, this gives a total of 1000 sample points, which is equivalent to a very sparse 6D sample grid of fewer than 4 points in each spacial translation, and an angular sampling interval of more than 90°. Such a sampling rate is too low for any competitive results, but about ten times slower than our FAML algorithm. The trade-off for computing fast alignment over more sample points is a highly efficient one.

In general, the scanning based accurate calculation of the integral in Equation 3 is in principle computationally infeasible, as the cost scales cubically with both the angular and translational sampling rates. For example, a sampling with a rotation angle interval of 2.5°, and a translational offset of ±10 voxels with a step of 1 voxel will take more than $7.4 \times 10^{10}$ sampled rigid transformations on the manifold $\Upsilon$.

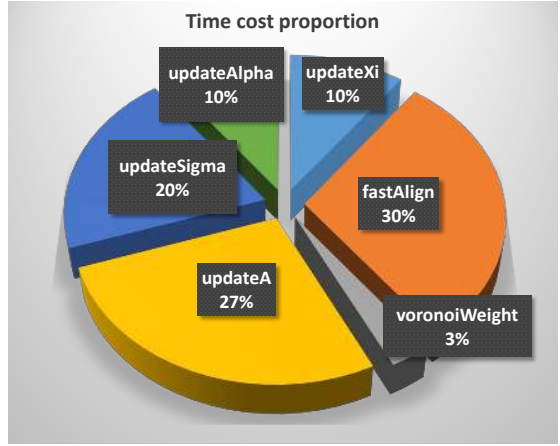

Figure 10: Pie chart of time cost proportions for major FAML steps for the averaging of subtomograms of size $128^3$ voxels in Table 1.

To our knowledge, in practice, adaptive oversampling similar to [28] has been used for approximating such integrals. In the later stages of the iterative averaging process, local orientation searches are used, which may degenerate such methods to the alignment based method by only sampling in the vicinity of a single rigid transformation between a subtomogram and a subtomogram average. The limits of alignment based subtomogram averaging methods may carry to such adaptive oversampling based maximum-likelihood subtomogram averaging methods. Theoretically, the multiple rigid transformations produced by our global search (Section 2.1) would prevent the averaging process from sticking at such local optima.

The computational time is not directly comparable because FA is implemented mainly in C++ and FAML (maximum-likelihood updating steps) is implemented in python. However, from a utility perspective, we compared the computation time used for FA and FAML averaging tasks. For each task, 20 simulated ribosome subtomograms (PDB ID: 4V4A) of specified sizes were averaged by both methods (the practical input subtomogram size limit is $256^3$ for current typical computer hardware settings). All simulated subtomograms are of SNR 0.02. Ribosome structures were randomly rotated and translated before they were used to construct the simulated subtomograms. We also include computation time cost using RELION (implemented in C++), the most popular subtomogram classification and averaging software which implements method [27]. RELION was tested using its default sampling parameters: 7.5° angular sampling interval without adaptive oversampling or local searches, and 5 pixels offset search range with 1 pixel search step. Using the default parameters gave us a rough estimation of the computation time cost. Note that in practice these sampling parameters should be modified accordingly with larger input subtomograms, which will further increase the computation time of the RELION method. All three methods were tested on the Intel Core I7 computer with 12 parallel computing processes.

Table 1: Computing time used for FA, FAML and RELION methods. $32^3$ in parenthesis denotes the testing subtomograms are of size $32^3$.

|  | Iterations to converge | Mean time per iteration | Total time |
| --- | --- | --- | --- |
| FA ($32^3$) | 11 | 7 s | 78 s |
| FAML ($32^3$) | 10 | 56 s | 562 s |
| RELION ($32^3$) | 8 | 340 s | 2720 s |
| FA ($64^3$) | 4 | 27 s | 106 s |
| FAML ($64^3$) | 3 | 150 s | 451 s |
| RELION ($64^3$) | 6 | 340 s | 2041 s |
| FA ($128^3$) | 5 | 143 s | 717 s |
| FAML ($128^3$) | 4 | 449 s | 1794 s |
| RELION ($128^3$) | 3 | 921 s | 2764 s |

We recorded in Table 1 the number of iterations it took to converge, time per iteration, and the total time it took to converge for each task. From Table 1, we found that though the FA method took less time per iteration, FAML generally took fewer iterations to converge. If the steps for maximum-likelihood update in FAML were implemented with C++, it would be expected to achieve several folds of speedup. Therefore, there is only a moderate increase of computation time of FAML compared to FA. Given that FA has achieved three magnitudes of speedup [32] compared to the orientation scanning exhaustive search based method [10], and that FAML requires a substantially smaller number of subtomograms with a faster convergence for successful structural recovery than FA does, we believe a moderate increase in time cost of FAML will not affect its efficacy for the systematic *de novel* recovery of large numbers of macromolecules with highly diverse structures and abundances captured by CECT.

## 4 Conclusion

CECT is a very promising tool for the systematic visualization of native structures and spatial organizations of large macromolecules inside single cells. Nevertheless, it remains one of the bottlenecks the efficient and accurate reference-free recovery and separation of large numbers of diverse macromolecular structures systematically through subtomogram averaging and classification. In this paper, building on existing work, we proposed a new method (FAML) that integrates fast subtomogram alignment [32] with maximum-likelihood [27] methods to improve the recognition and recovery of initial structural models from input subtomograms. Our experiments showed a significant improvement compared with our previous methods [32, 15] in terms of 1) the number of subtomograms needed for successful recovery and classification, and 2) robustness to the noise and missing wedge effects. FAML is favored especially with subtomogram datasets of low SNR or tilt angle range.

Due to its high scalability and accuracy, FAML is a very useful component for improved systematic structural pattern mining in CECT, thereby bridging the gap from microscopy to structure. A potential use of FAML is to combine it with other reference-free structural pattern mining techniques. For example, given millions of subtomograms extracted from cellular tomograms using reference-free particle picking (e.g. [29]), these macromolecules can be first filtered using our recently developed deep learning based coarse structural separation method (Section 3.2.2) [33], then be classified and averaged using FAML. The resulting averages can be further refined by maximum-likelihood methods that take into account Contrast Transfer Functions [7] or high-precision alignment [31] method. Besides CECT, FAML can be applied to similar data analysis tasks from cryo-tomograms of purified complexes or cell lysate.

## Acknowledgements

We thank Dr. Robert F. Murphy for suggestions. We thank Dr. Zachary Freyberg and Dr. James Krieger for technical assistance. We thank Dr. Achilleas Frangakis and Dr. Michael Kunz for sharing the TMV subtomograms for the averaging test. We thank Dr. Friedrich Förster for sharing the GroEL and GroEL-ES subtomograms for the classification test.

## Funding

This work was supported in part by U.S. National Institutes of Health (NIH) grant P41 GM103712. M.X acknowledges support of Samuel and Emma Winters Foundation.